\documentstyle[12pt,thmsa,sw20lart]{article}

\input tcilatex
\QQQ{Language}{
American English
}

\begin{document}

\title{Retardation Terms in The One-Gluon Exchange Potential }
\author{Jun-Chen Su and Jian-Xing Chen \\
Center for Theoretical Physics, Department of Physics,\\
Jilin University, Changchun 130023,\\
People's Republic of China}
\date{}
\maketitle

\begin{abstract}
It is pointed out that the retardation terms given in the original
Fermi-Breit potential vanish in the center of mass frame. The retarded
one-gluon exchange potential is rederived in this paper from the
three-dimensional one-gluon exchange kernel which appears in the exact
three-dimensional relativistic equation for quark-antiquark bound states.
The retardation part of the potential given in the approximation of order $%
p^2/m^2$ is shown to be different from those derived in the previous
literature. This part is off-shell and does no longer vanish in the center
of mass frame.

PACS number(s): 12.40 Qq, 11.10.St, 14.40.Gx.

\end{abstract}

In the quark potential model$^{\left[ 1-6\right] }$, the one-gluon exchange
potential (OGEP), or say, the Fermi-Breit (F-B) potential, usually is chosen
to describe the short-range interaction between two quarks ($qq$), two
antiquarks ($\overline{q}\overline{q}$) or a quark and an antiquark ($q%
\overline{q}$). This potential was originally derived from the t-channel
one-gluon exchange S-matrix$^{\left[ 7\right] }$ or the t-channel one-gluon
exchange kernel appearing in the Bethe-Salpeter (B-S) equation$^{\left[
4,5\right] }$. In the relativistic version, the potential may be represented
in the form 
\begin{equation}
V(p_1^{^{\prime }},p_2^{^{\prime }};p_1,p_2)=U^{+}(\overrightarrow{p}%
_1^{^{\prime }})U^{+}(\overrightarrow{p}_2^{^{\prime }})\widehat{V}(1,2)U(%
\overrightarrow{p}_1)U(\overrightarrow{p}_2)
\end{equation}
where 
\begin{equation}
U(\overrightarrow{p}_i)=\sqrt{\frac{E_i+m_i}{2E_i}}\binom 1{\frac{%
\overrightarrow{\sigma }_i\cdot \overrightarrow{p}_i}{E_i+m_i}}
\end{equation}
is the Dirac spinor for ith quark, satisfying the normalization condition $%
U^{+}(\overrightarrow{p}_i)$ $U(\overrightarrow{p}_i)=1$ and 
\begin{eqnarray}
\widehat{V}(1,2) &=&i\gamma _1^0\gamma _2^0K(1,2) \\
\ &=&g^2\widehat{C}_s\gamma _1^0\gamma _1^\mu D_{\mu \nu }(Q)\gamma
_2^0\gamma _2^\nu  \nonumber
\end{eqnarray}
is the interaction operator. The first equality in Eq.(3) shows the general
relation of the interaction operator with the B-S irreducible kernel K(1,2),
while the second equality gives the expression for the t-channel one-gluon
exchange interaction in which $\widehat{C}_s$ stands for the color matrix 
\begin{equation}
\widehat{C}_s=\left\{ 
\begin{array}{l}
\frac{\lambda _1^c}2\frac{\lambda _2^c}2(\frac{\lambda _1^{c^{*}}}2\frac{%
\lambda _2^{c^{*}}}2),for\text{ }qq\text{ }(\overline{q}\overline{q}), \\ 
-\frac{\lambda _1^c}2\frac{\lambda _2^{c^{*}}}2,\text{ }for\text{ }q%
\overline{q}.
\end{array}
\right.
\end{equation}
and $D_{\mu \nu }(Q)$ denotes the gluon propagator whose argument $Q$ is
defined as 
\begin{equation}
Q=p_1^{^{\prime }}-p_1=p_2-p_2^{^{\prime }}
\end{equation}

Upon inserting Eq.(3) into Eq.(1), in the Feynman gauge which was often used
in the previous derivation of the FBP, we may write 
\begin{equation}
V(p_1^{^{\prime }},p_2^{^{\prime }};p_1,p_2)=4\pi \alpha _s\widehat{C}%
_s\Gamma (\overrightarrow{p}_1^{^{\prime }},\overrightarrow{p}_2^{^{\prime
}};\overrightarrow{p}_1,\overrightarrow{p}_2)D(Q)
\end{equation}
where $\alpha _s=g^2/4\pi $, 
\begin{equation}
\Gamma (\overrightarrow{p}_1^{^{\prime }},\overrightarrow{p}_2^{^{\prime }};%
\overrightarrow{p}_1,\overrightarrow{p}_2)=U^{+}(\overrightarrow{p}%
_1^{^{\prime }})\gamma _1^0\gamma _1^\mu U(\overrightarrow{p}_1)U^{+}(%
\overrightarrow{p}_2^{^{\prime }})\gamma _2^0\gamma _{2\mu }U(%
\overrightarrow{p}_2)
\end{equation}
and 
\begin{equation}
D(Q)=-\frac 1{Q^2+i\epsilon }
\end{equation}
Ordinarily, the OGEP is derived in the nonrelativistic approximation up to
the order of $p^2/m^2(v^2/c^2)$ and given in the three-dimensional space. By
making use of the nonrelativistic expression for the quark energy 
\begin{equation}
E_i=\sqrt{\overrightarrow{p_i}^2+m_i^2}\approx m_i+\frac{\overrightarrow{p_i}%
^2}{2m_i}
\end{equation}
the three-dimensional spinor matrix in Eq.(7) will be written, up to the
order $p^2/m^2$ , as follows$^{\left[ 1,7\right] }$%
\begin{eqnarray}
\Gamma (\overrightarrow{P},\overrightarrow{q},\overrightarrow{k}) &=&1-\frac{%
\overrightarrow{P}^2}{m_{12}^2}-\frac{(m_1-m_2)^2}{8m_1^2m_2^2}(%
\overrightarrow{q}-\overrightarrow{k})^2- \\
&&\ \ \frac{m_1-m_2}{2m_1m_2m_{12}}\ \overrightarrow{P}\cdot (%
\overrightarrow{q}+\overrightarrow{k})+\frac{\overrightarrow{q\cdot }%
\overrightarrow{k}}{m_1m_2}  \nonumber \\
&&\ \ +\frac i{4m_{12}}[\overrightarrow{P}\times (\overrightarrow{q}-%
\overrightarrow{k})]\cdot (\frac{\overrightarrow{\sigma }_1}{m_1}-\frac{%
\overrightarrow{\sigma }_2}{m_2})  \nonumber \\
&&\ \ \ -\frac{(\overrightarrow{q}-\overrightarrow{k})^2}{6m_1m_2}%
\overrightarrow{\sigma }_1\cdot \overrightarrow{\sigma }_2+\frac{(%
\overrightarrow{q}-\overrightarrow{k})^2}{4m_1m_2}T_{12}(\overrightarrow{q}-%
\overrightarrow{k})  \nonumber \\
&&\ \ \ +\frac i{4m_1m_2}(\overrightarrow{q}\times \overrightarrow{k})\cdot
[(2+\frac{m_2}{m_1})\overrightarrow{\sigma }_1+(2+\frac{m_1}{m_2})%
\overrightarrow{\sigma }_2]  \nonumber
\end{eqnarray}
where 
\begin{equation}
T_{12}(\overrightarrow{q}-\overrightarrow{k})=\frac 1{\left| \overrightarrow{%
q}-\overrightarrow{k}\right| ^2}\overrightarrow{\sigma }_1\cdot (%
\overrightarrow{q}-\overrightarrow{k})\overrightarrow{\sigma }_2\cdot (%
\overrightarrow{q}-\overrightarrow{k})-\frac 13\overrightarrow{\sigma }%
_1\cdot \overrightarrow{\sigma }_2
\end{equation}
is the tensor force. In the above, we have introduced the total momentum $%
\overrightarrow{P}$ and the relative momentum $\overrightarrow{k}$ and $%
\overrightarrow{q}$ for the initial and final states of the two particles
respectively. They are defined by 
\begin{eqnarray}
\overrightarrow{p}_1 &=&\eta _1\overrightarrow{P}+\overrightarrow{k},%
\overrightarrow{p}_2=\eta _2\overrightarrow{P}-\overrightarrow{k} \\
\overrightarrow{p}_1^{^{\prime }} &=&\eta _1\overrightarrow{P}+%
\overrightarrow{q},\overrightarrow{p}_2^{^{\prime }}=\eta _2\overrightarrow{P%
}-\overrightarrow{q}  \nonumber \\
\eta _i &=&\frac{m_i}{m_{12}},m_{12}=m_1+m_2,i=1,2  \nonumber
\end{eqnarray}
To obtain a three-dimensional potential, the question left is centered on
how to reduce the four-dimensional gluon propagator to a three-dimensional
form. In the original derivation of the F-B potential, with the recognition
that the Coulomb term in the potential is dominant$^{\left[ 2,6\right] }$,
meaning that the ratio $Q_0^2/\overrightarrow{Q}^2$ is much less than unity,
the vector boson propagator was reasonably expanded as$^{\left[ 7\right] }$%
\begin{equation}
D(Q)=\frac 1{\overrightarrow{Q}^2-Q_0^2}\approx \frac 1{\overrightarrow{Q}%
^2}(1+\frac{Q_0^2}{\overrightarrow{Q}^2})
\end{equation}
Based on the energy conservation denoted in Eq.(5) and the nonrelativistic
fermion energies written in Eq.(9), in the approximation of the order $%
p^2/m^2$ , the time-component of the transfer momentum squared in Eq.(13)
may be taken to be $^{\left[ 7\right] }$%
\begin{eqnarray}
Q_0^2 &=&(E_1^{^{\prime }}-E_1)(E_2-E_2^{^{\prime }}) \\
\ &\approx &\frac 1{4m_1m_2}(4\overrightarrow{Q}\cdot \overrightarrow{p}_1%
\overrightarrow{Q}\cdot \overrightarrow{p}_2+2\overrightarrow{Q}^2%
\overrightarrow{Q}\cdot \overrightarrow{p}_2-2\overrightarrow{Q}^2%
\overrightarrow{Q}\cdot \overrightarrow{p}_1-\overrightarrow{Q}^4)  \nonumber
\end{eqnarray}
On substituting this expression into Eq.(13), the propagator was
approximately represented in the following three-dimensional form$^{\left[
1,7\right] }$%
\begin{equation}
D(Q)=\frac 1{\overrightarrow{Q}^2}-\frac 1{4m_1m_2}-\frac{\overrightarrow{Q}%
\cdot (\overrightarrow{p}_1-\overrightarrow{p}_2)}{2m_1m_2\overrightarrow{Q}%
^2}+\frac{\overrightarrow{Q}\cdot \overrightarrow{p}_1\overrightarrow{Q}%
\cdot \overrightarrow{p}_2}{m_1m_2\overrightarrow{Q}^4}
\end{equation}
where the first term gives the coulombic potential which describes the
instantaneous interaction, while the remaining terms represent the
retardation effect. When Eqs.(10) and (15) are inserted into Eq.(6) and the
terms higher than $p^2/m^2$ are neglected, the well-known F-B potential will
be recovered.

It should be pointed out that according to the procedure of derivation
stated above, the retardation terms in the F-B potential may possibly be
given different expressions. This is because the $Q_0^2$ In Eq.(13) may be
represented in different ways. For example, in accordance with the relation
shown in Eq.(5), we may also write $Q_0^2=(E_1^{^{\prime }}-E_1)^2$ or $%
Q_0^2=(E_2-E_2^{^{\prime }})^2$. Corresponding to either of these two
representations, in the approximation of order $p^2/m^2$, the propagator
will be given a different expression 
\begin{equation}
D(Q)=\frac 1{\overrightarrow{Q}^2}+\frac 1{4m_1^2}+\frac{(\overrightarrow{Q}%
\cdot \overrightarrow{p}_1)^2}{m_1^2\overrightarrow{Q}^4}+\frac{%
\overrightarrow{Q}\cdot \overrightarrow{p}_1}{m_1^2\overrightarrow{Q}^2}
\end{equation}
or

\begin{equation}
D(Q)=\frac 1{\overrightarrow{Q}^2}+\frac 1{4m_2^2}+\frac{(\overrightarrow{Q}%
\cdot \overrightarrow{p}_2)^2}{m_2^2\overrightarrow{Q}^4}-\frac{%
\overrightarrow{Q}\cdot \overrightarrow{p}_2}{m_2^2\overrightarrow{Q}^2}
\end{equation}
The different expressions given in Eqs.(15)-(17) will be consistent with
each other when and only when they are constrained by the energy-momentum
conservation. In fact, when the nonrelativistic energies in Eq.(9) are
substituted into the energy-conservation 
\begin{equation}
E_1^{^{\prime }}+E_2^{^{\prime }}=E_1+E_2
\end{equation}
and then the momentum conservation shown in Eq.(5) is employed, one may find 
\begin{equation}
\overrightarrow{Q}^2=\frac 2{m_{12}}(m_1\overrightarrow{p}_2-m_2%
\overrightarrow{p}_1)\cdot \overrightarrow{Q}
\end{equation}
With this relation, it is easy to prove that the expressions in
Eqs.(15)-(17) may be transformed to one another. This proof indicates that
the energy-momentum conservation acts a necessary constraint to guarantee
the retardation terms in the F-B potential to be uniquely determined.

When the energies in Eq.(18) are replaced by the nonrelativistic ones shown
in Eq.(9) and then the momenta of single particles are represented through
the total and relative momenta defined in Eq.(12) , it will be found 
\begin{equation}
\overrightarrow{q}^2=\overrightarrow{k}^2
\end{equation}
With the aid of this relation and those shown in Eq.(12), the $Q_0^2$
defined in the first line of Eq.(14) can be rewritten as 
\begin{eqnarray}
Q_0^2 &=&\frac 1{4m_1m_2}(\overrightarrow{p}_1^{^{\prime }2}-\overrightarrow{%
p}_1^2)(\overrightarrow{p}_2^2-\overrightarrow{p}_2^{^{\prime }2}) \\
&=&\frac 1{m_{12}^2}(\overrightarrow{Q}\cdot \overrightarrow{P})^2  \nonumber
\end{eqnarray}
This result may directly be derived from the last expression of Eq.(14),
when the momenta $\overrightarrow{p}_1$ and $\overrightarrow{p}_2$ in the
expression are replaced by the total and relative momenta and then using the
following relation 
\begin{equation}
\overrightarrow{Q}\cdot \overrightarrow{k}=-\frac 12\overrightarrow{Q}^2
\end{equation}
which is obtained from Eq.(19) by employing the relations denoted in Eq.(12)$%
.$ From Eq.(21) , it is clear to see that in the center of mass frame ($%
\overrightarrow{P}=0$ ), we have 
\begin{equation}
Q_0^2=0
\end{equation}
Due to the above equality, Eq.(13) becomes 
\begin{equation}
D(Q)=\frac 1{\overrightarrow{Q}^2}
\end{equation}
which is completely instantaneous. This result shows that the ordinary F-B
potential fails to describe the retardation effect existing in the center of
mass frame. The failure obviously originates from the use of the
on-mass-shell approximation shown in Eq.(14) in the derivation of the
three-dimensional approximate propagator given in Eq.(15). The above result
also indicates that only the instantaneous part of the F-B potential can be
reasonably used in the study of hadron spectroscopy.

Quite recently, the retardation effect has attracted some interest in the
study of hadron spectroscopy$^{\left[ 6,8,9\right] }$. In Ref.(8), the
authors derived a certain retardation terms for the scalar and vector
couplings in a semirelativistic quark potential model. In Ref.(9) , the
retardation effect was especially investigated in some detail. In the
investigation, the authors did not give an explicit three-dimensional
expression of the retardation corrections to the potential. Instead, they
directly evaluated the retardation corrections to the quarkonium with the
help of a regularization procedure. The retardation corrections were defined
as the difference between the values obtained from solving the four
dimensional B-S equation for bound states by using the exact one gluon
exchange B-S kernel and the instantaneous one.

In this paper, we study the retardation effect from a different angle and
are devoted to deriving explicit expressions of the retardation part of the
OGEP. The essential feature of the derivation is that we start from an exact
three-dimensional relativistic equation for $q\overline{q}$($qq$ or $%
\overline{q}\overline{q}$ ) bound states which is equivalent to the
corresponding four-dimensional B-S equation$^{\left[ 10,11\right] }$. 
\begin{eqnarray}
&&\ \ [E-h(\overrightarrow{p}_1)-h(\overrightarrow{p}_2)]\chi _{p\alpha }(%
\overrightarrow{q}) \\
\  &=&\int \frac{d^3k}{\left( 2\pi \right) ^3}K(\overrightarrow{p},%
\overrightarrow{q},\overrightarrow{k},E)\chi _{p\alpha }(\overrightarrow{k})
\nonumber
\end{eqnarray}
where $E$ is the total energy of the system, $\chi _{p\alpha }(%
\overrightarrow{q})$ is the wave function for the system, 
\begin{equation}
h(\overrightarrow{p}_i)=\overrightarrow{\alpha }_i\cdot \overrightarrow{p_i}%
+\beta _im_i,\text{ }i=1,2
\end{equation}
is the Hamiltonian for the ith fermion and $K(\overrightarrow{p},%
\overrightarrow{q},\overrightarrow{k},E)$ denotes the three-dimensional
interaction kernel which may be explicitly expressed in terms of a few types
of Green's functions and is energy-dependent as any interaction kernel$%
^{[10,11]}$. In this paper, we only concern the t-channel one-gluon exchange
kernel. This kernel still has the form as represented in Eq.(3) except that
the gluon propagator becomes three-dimensional. In the Feynman gauge, the
propagator is of the form$^{\left[ 10,11\right] }$%
\begin{eqnarray}
D(\overrightarrow{p},\overrightarrow{q},\overrightarrow{k},E) &=&-\frac
1{2Q}[\frac 1{E-E(\overrightarrow{p}_2)-E(\overrightarrow{p}_1^{^{\prime
}})-Q+i\epsilon } \\
&&\ \ +\frac 1{E-E(\overrightarrow{p}_1)-E(\overrightarrow{p}_2^{^{\prime
}})-Q+i\epsilon }]  \nonumber
\end{eqnarray}
where $Q=\left| \overrightarrow{q}-\overrightarrow{k}\right| $ is the
magnitude of the three-dimensional transfer momentum, E is the total energy
of the interacting system,  $E(\overrightarrow{p}_1)$, $E(\overrightarrow{p}%
_2)$, $E(\overrightarrow{p}_1^{^{\prime }})$ and $E(\overrightarrow{p}%
_2^{^{\prime }})$ are the energies of single free particles which are on the
mass-shell.  It is easy to verify that when the total energy in Eqs.(8) and
(27) is put on the mass shell, $E=E(\overrightarrow{p}_1)+$ $E(%
\overrightarrow{p}_2)=$ $E(\overrightarrow{p}_1^{^{\prime }})$ $+$ $E(%
\overrightarrow{p}_2^{^{\prime }}),$  the both expressions in Eqs.(8) and
(27) may be transformed to each other. When the energy $E$ is off-shell, 
the propagator shown above just is the exact three-dimensional
representation of the propagator written in Eq.(8) and therefore containing
all the retardation effects in the case of bound states. 

Now let us discuss the retardation terms in the OGEP. It is clear that when
the propagator in Eq.(27) is inserted into Eq.(6), we obtain a rigorous
relativistic representation of the retarded OGEP written in the
three-dimensional space. In the following, we are interested in deriving a
retarded OGEP in the nonrelativistic approximation which is used in the
nonrelativistic quark potential model. For this purpose, it is necessary to
write the propagator in Eq.(27) in the nonrelativistic approximation of the
order $p^2/m^2$ by employing the approximate expression in Eq.(9) 
\begin{eqnarray}
D(\overrightarrow{P},\overrightarrow{q},\overrightarrow{k}) &=&\frac
1{2Q}[\frac 1{Q-\varepsilon +\frac{\overrightarrow{p}_1^{^{\prime }2}}{2m_1}+%
\frac{\overrightarrow{p}_2^2}{2m_2}-i\epsilon } \\
&&\ +\frac 1{Q-\varepsilon +\frac{\overrightarrow{p}_2^{^{\prime }2}}{2m_2}+%
\frac{\overrightarrow{p}_1^2}{2m_1}-i\epsilon }]  \nonumber
\end{eqnarray}
where $\varepsilon =E-m_1-m_2$ is the binding energy. Firstly, we discuss a
special case that the binding energy is small, being of the order of $%
p^2/m^2.$ In this case, noticing the common recognition mentioned before for
the expansion shown in Eq.(13) that in the nonrelativistic approximation,
the Coulomb term plays a dominant role in the OGEP$^{\left[ 2,6,7\right] },$
the above propagator may approximately be taken to be$^{\left[ 11\right] }$%
\begin{equation}
D(\overrightarrow{P},\overrightarrow{q},\overrightarrow{k})=\frac
1{Q^2}+\frac 1{Q^3}[\varepsilon -\frac{\overrightarrow{P}^2}{2m_{12}}-\frac
1{4\mu }(\overrightarrow{q}^2+\overrightarrow{k}^2)]
\end{equation}
where $\mu $ is the reduced mass of two particles, $\overrightarrow{P},%
\overrightarrow{q}$and$\overrightarrow{k}$ are the momenta defined in
Eq.(12). In Eq.(29), except for the first term which is instantaneous, the
other terms all come from the retardation effect. In contrast to the
corresponding expression given in Eq.(15), the propagator in Eq.(29) is
off-shell, i.e., energy-dependent, and, as we see, the retardation terms in
it no longer vanish in the center of mass frame. Substituting Eqs$.$(10) and
(29) into Eq.(6), the OGEP may be separated into two parts: the
instantaneous part $V_I(\overrightarrow{P},\overrightarrow{q},%
\overrightarrow{k})$ and the retardation part $V_R(\overrightarrow{P},%
\overrightarrow{q},\overrightarrow{k})$%
\begin{equation}
V(\overrightarrow{P},\overrightarrow{q},\overrightarrow{k})=V_I(%
\overrightarrow{P},\overrightarrow{q},\overrightarrow{k})+V_R(%
\overrightarrow{P},\overrightarrow{q},\overrightarrow{k})
\end{equation}
where 
\begin{equation}
V_I(\overrightarrow{P},\overrightarrow{q},\overrightarrow{k})=\frac{4\pi
\alpha _s\widehat{C}_s}{\left| \overrightarrow{q}-\overrightarrow{k}\right|
^2}\Gamma (\overrightarrow{P},\overrightarrow{q},\overrightarrow{k})
\end{equation}
in which the $\Gamma (\overrightarrow{P},\overrightarrow{q},\overrightarrow{k%
})$ was represented in Eq.(10) and 
\begin{equation}
V_R(\overrightarrow{P},\overrightarrow{q},\overrightarrow{k})=\frac{4\pi
\alpha _s\widehat{C}_s}{\left| \overrightarrow{q}-\overrightarrow{k}\right|
^3}[\varepsilon -\frac{\overrightarrow{P}^2}{2m_{12}}-\frac 1{4\mu }(%
\overrightarrow{q}^2+\overrightarrow{k}^2)]
\end{equation}
By the Fourier transformation 
\begin{eqnarray}
V(\overrightarrow{P},\overrightarrow{r},\overrightarrow{r}^{^{\prime }})
&=&\int \frac{d^3q}{\left( 2\pi \right) ^3}\frac{d^3k}{\left( 2\pi \right) ^3%
}e^{i\overrightarrow{q}\cdot \overrightarrow{r}-i\overrightarrow{k}\cdot 
\overrightarrow{r}^{^{\prime }}}V(\overrightarrow{P},\overrightarrow{q},%
\overrightarrow{k}) \\
\ &=&\delta ^3(\overrightarrow{r}-\overrightarrow{r}^{^{\prime }})V(%
\overrightarrow{P},\overrightarrow{r})  \nonumber
\end{eqnarray}
We may obtain a local OGEP in the position space 
\begin{equation}
V(\overrightarrow{P},\overrightarrow{r})=V_I(\overrightarrow{P},%
\overrightarrow{r})+V_R(\overrightarrow{P},\overrightarrow{r})
\end{equation}
where $V_I(\overrightarrow{P},\overrightarrow{r})$ and $V_R(\overrightarrow{P%
},\overrightarrow{r})$ are respectively the instantaneous part and the
retardation part of the potential $V(\overrightarrow{P},\overrightarrow{r})$%
. The instantaneous part is$^{\left[ 1,11\right] }$%
\begin{eqnarray}
V_I(\overrightarrow{P},\overrightarrow{r}) &=&\frac{\alpha _s\widehat{C}_s}%
r\{1-\frac{\overrightarrow{P}^2}{m_{12}}+\frac{i\left( m_1-m_2\right) }{%
2m_1m_2m_{12}r^2}(\overrightarrow{r}\cdot \overrightarrow{P} \\
&&\ -2r^2\overrightarrow{P}\cdot \nabla _{\overrightarrow{r}})-\frac{\pi
\left( m_1-m_2\right) ^2}{2m_1^2m_2^2}r\delta ^3(\overrightarrow{r})-\frac
1{m_1m_2r^2}(r^2\nabla _{\overrightarrow{r}}^2  \nonumber \\
&&\ -\overrightarrow{r}\cdot \nabla _{\overrightarrow{r}})-\frac{8\pi }{%
3m_1m_2}r\delta ^3(\overrightarrow{r})\overrightarrow{S_1}\cdot 
\overrightarrow{S_2}-\frac 3{4m_1m_2r^2}T(\overrightarrow{r})  \nonumber \\
&&\ +\frac 1{2m_{12}r^2}\overrightarrow{L}_R\cdot (\frac{\overrightarrow{S_1}%
}{m_1}-\frac{\overrightarrow{S_2}}{m_2})  \nonumber \\
&&\ -\frac 1{2m_1m_2r^2}\overrightarrow{L}_r\cdot [(2+\frac{m_2}{m_1})%
\overrightarrow{S_1}+(2+\frac{m_1}{m_2})\overrightarrow{S_2}]\}  \nonumber
\end{eqnarray}
where 
\begin{eqnarray}
\overrightarrow{P} &=&-i\nabla _{\overrightarrow{R}},\overrightarrow{S_i}=%
\frac{\overrightarrow{\sigma }_i}2,i=1,2 \\
\overrightarrow{L}_R &=&\overrightarrow{r}\times (-i\nabla _{\overrightarrow{%
R}}),\overrightarrow{L}_r=\overrightarrow{r}\times (-i\nabla _{%
\overrightarrow{r}})  \nonumber \\
T(\overrightarrow{r}) &=&\frac 1{r^2}\overrightarrow{\sigma }_1\cdot 
\overrightarrow{r}\overrightarrow{\sigma }_2\cdot \overrightarrow{r}-\frac 13%
\overrightarrow{\sigma }_1\cdot \overrightarrow{\sigma }_2  \nonumber
\end{eqnarray}
$\overrightarrow{R}$ and $\overrightarrow{r}$ denote the center of mass
coordinate and the relative coordinate respectively. For the retardation
part, we have$^{\left[ 11\right] }$%
\begin{equation}
V_R(\overrightarrow{P},\overrightarrow{r})=-\frac 2\pi \alpha _s\widehat{C}%
_s\{\varepsilon \text{ }\ln \frac r{r_0}-\frac{\overrightarrow{P}^2}{2m_{12}}%
+\frac 1{4\mu r^2}[1+2\overrightarrow{r}\cdot \nabla _{\overrightarrow{r}%
}+2r^2\ln (\frac r{r_0})\nabla _{\overrightarrow{r}}^2]\}
\end{equation}
where the Fourier transformation 
\begin{equation}
\int \frac{d^3Q}{\left( 2\pi \right) ^3}\frac{e^{i\overrightarrow{Q}\cdot 
\overrightarrow{r}}}{\left| \overrightarrow{Q}\right| ^3}=-\frac 1{2\pi ^2}%
\text{ }\ln \frac r{r_0}
\end{equation}
in which r$_0$ is a size parameter has been used. The logarithmic function
in Eq.(37) was ever phenomenologically introduced to the quark potential
model in some previous works$^{\left[ 12\right] }$ for describing the
intermediate-range interaction. Now, it appears as a natural result of the
retardation effect.

Next, we have a particular interest in deriving a retarded OGEP in a more
general case that the binding energy is not required to be small. In this
case, the gluon propagator in Eq.(28) should be expanded in the form in the
approximation of order $p^2/m^2$ as follows 
\begin{eqnarray}
D(\overrightarrow{P},\overrightarrow{q},\overrightarrow{k}) &=&\frac
1{Q(Q-\varepsilon )}\{1-\frac 1{Q-\varepsilon }[\frac{\overrightarrow{P}^2}{%
2m_{12}}+\frac 1{4\mu }(\overrightarrow{q}^2+\overrightarrow{k}^2)]\} \\
\ &=&\frac 1{Q^2}+D_R(\overrightarrow{P},\overrightarrow{q},\overrightarrow{k%
})  \nonumber
\end{eqnarray}
where 
\begin{equation}
D_R(\overrightarrow{P},\overrightarrow{q},\overrightarrow{k})=D_1(Q)-D_2(Q)[%
\frac{\overrightarrow{P}^2}{2m_{12}}+\frac 1{4\mu }(\overrightarrow{q}^2+%
\overrightarrow{k}^2)]
\end{equation}
in which 
\begin{equation}
D_1(Q)=\frac \varepsilon {Q^2(Q-\varepsilon )}
\end{equation}
and 
\begin{equation}
D_2(Q)=\frac 1{Q(Q-\varepsilon )^2}
\end{equation}
With the propagator given in Eq.(39), the total OGEP may also be divided
into the two parts as written in Eq.(30). The instantaneous part of the
potential is still the one represented in Eq.(31) in the momentum space and
the one in Eq.(35) in the position space. While, the retardation part of the
potential now is of the following form in the approximation of order $%
p^2/m^2 $%
\begin{equation}
V_R(\overrightarrow{P},\overrightarrow{q},\overrightarrow{k})=4\pi \alpha _s%
\widehat{C}_s\{D_1(Q)\Gamma (\overrightarrow{P},\overrightarrow{q},%
\overrightarrow{k})-D_2(Q)[\frac{\overrightarrow{P}^2}{2m_{12}}+\frac 1{4\mu
}(\overrightarrow{q}^2+\overrightarrow{k}^2)]\}
\end{equation}
Through the Fourier transformation as written in Eq.(33) , we may obtain
from Eq.(43) a local form of the retardation part of the potential as
follows 
\begin{eqnarray}
V_R(\overrightarrow{P},\overrightarrow{r}) &=&4\pi \alpha _s\widehat{C}%
_s\{[1-\frac{\overrightarrow{P}^2}{m_{12}^2}+\frac{(m_1-m_2)^2}{8m_1^2m_2^2}%
\nabla _{\overrightarrow{r}}^2+\frac 1{2m_{12}}(\overrightarrow{P}\times
\nabla _{\overrightarrow{r}})  \nonumber \\
&&\cdot (\frac{\overrightarrow{S_1}}{m_1}-\frac{\overrightarrow{S_2}}{m_2}%
)+\frac 1{m_1m_2}\overrightarrow{S_1}\cdot \overrightarrow{S_2}\nabla _{%
\overrightarrow{r}}^2-\frac 1{m_1m_2}\overrightarrow{S_1}\cdot \nabla _{%
\overrightarrow{r}}\ \overrightarrow{S_2}\cdot \nabla _{\overrightarrow{r}}]
\nonumber \\
&&\times D_1(r)-\frac{i\left( m_1-m_2\right) }{2m_1m_2m_{12}}\overrightarrow{%
P}\cdot [\nabla _{\overrightarrow{r}}D_1(r)+2D_1(r)\nabla _{\overrightarrow{r%
}}] \\
&&\ -\frac 1{m_1m_2}[\nabla _{\overrightarrow{r}}D_1(r)\cdot \nabla _{%
\overrightarrow{r}}+D_1(r)\nabla _{\overrightarrow{r}}^2]-\frac
i{2m_1m_2}[\nabla _{\overrightarrow{r}}D_1(r)  \nonumber \\
&&\times \ \nabla _{\overrightarrow{r}}]\cdot [(2+\frac{m_2}{m_1})%
\overrightarrow{S_1}+(2+\frac{m_1}{m_2}\overrightarrow{)S_2}]-\frac{%
\overrightarrow{P}^2}{2m_{12}}D_2(r)  \nonumber \\
&&\ +\frac 1{4\mu }[\nabla _{\overrightarrow{r}}^2D_2(r)+2\nabla _{%
\overrightarrow{r}}D_2(r)\cdot \nabla _{\overrightarrow{r}}+2D_2(r)\nabla _{%
\overrightarrow{r}}^2]\}  \nonumber
\end{eqnarray}
where 
\begin{eqnarray}
D_1(r) &=&\int \frac{d^3Q}{(2\pi )^3}\frac{\varepsilon \text{ }e^{i%
\overrightarrow{Q}\cdot \overrightarrow{r}}}{\overrightarrow{Q}^2(\left| 
\overrightarrow{Q}\right| -\varepsilon )} \\
\ &=&\frac 1{4\pi r}\cos x_0-\frac \varepsilon {2\pi ^2}\stackrel{}{%
\stackrel{\infty }{\stackunder{0}{\int }}\frac{dx\text{ }e^{-x}}{x^2+x_0^2}}
\nonumber \\
\ &=&\frac 1{4\pi r}\{\cos x_0-1+\frac 2\pi \ln x_0\sin x_0-\frac 2\pi [\xi
(x_0)\cos x_0-\eta (x_0)\sin x_0]\}  \nonumber
\end{eqnarray}
in which $x_0=\left| \varepsilon \right| r$, 
\begin{equation}
\xi (x_0)=\stackunder{n=0}{\stackrel{\infty }{\sum }}\frac{\left( -1\right)
^nx_0^{2n+1}}{(2n+1)(2n+1)!},\eta (x_0)=1+\stackunder{n=1}{\stackrel{\infty 
}{\sum }}\frac{\left( -1\right) ^nx_0^{2n}}{2n(2n)!}
\end{equation}
and 
\begin{eqnarray}
D_2(r) &=&\int \frac{d^3Q}{(2\pi )^3}\frac{\text{ }e^{i\overrightarrow{Q}%
\cdot \overrightarrow{r}}}{\left| \overrightarrow{Q}\right| (\left| 
\overrightarrow{Q}\right| -\varepsilon )^2}  \nonumber \\
\ &=&\frac 1{2\pi ^2}\stackrel{\infty }{\stackunder{0}{\int }}dx\frac{\sin x%
}{(x+x_0)^2} \\
\ &=&\frac 1{2\pi ^2}[\frac \pi 2\sin x_0-\ln x_0\cos x_0-\xi (x_0)\sin
x_0-\eta (x_0)\cos x_0]  \nonumber
\end{eqnarray}
In the practical calculation, the integral expressions in Eqs.(45) and (47)
are more suitable to be used. In the case of the binding energy being small,
the functions $D_1(r$ $)$ and $D_2(r)$ will be approximated to 
\begin{equation}
D_1(r)\approx -\frac \varepsilon {2\pi ^2}\ln (\left| \varepsilon \right| r)
\end{equation}
and 
\begin{equation}
D_2(r)\approx -\frac 1{2\pi ^2}[1+\ln (\left| \varepsilon \right| r)]
\end{equation}
With this approximation, the expression in Eq.(44) will be reduced to the
one given in Eq.(37) if we set $\left| \varepsilon \right| =r_0^{-1}$ in the
function $\ln (\left| \varepsilon \right| r)$.

\subsection{Acknowledgments}

The authors would like to thank Professor Shi-Shu Wu for useful discussions.
This work was supported in part by National Natural Science Foundation of
China and in part by The Research Fund for the Doctoral Program of Higher
Education.

\subsection{References}

\begin{description}
\item[1]  A. De Rujula, H. Georgi and S. L. Glashow, Phys. Rev. D12, 147
(1975).
\end{description}

\begin{quote}
.
\end{quote}

\end{document}